# *Towards automation of the polyol process for the synthesis of silver nanoparticles*

*Jakob Wolf[1,2], Tomasz M. Stawski[1]\*, Glen J. Smales[1], Andreas F. Thünemann[1], Franziska Emmerling[1,3]\*\**

\* tomasz.stawski@bam.de; \*\* franziska.emmerling@bam.de

[1]Federal Institute for Materials Research and Testing (BAM), 12489, Berlin, Germany;

[2]Max Planck Institute of Colloids and Interfaces, 14476 Potsdam, Germany;

[3]Humboldt-Universität zu Berlin, Department of Chemistry, 12489, Berlin, Germany

ORCID:

JW: https://orcid.org/0000-0003-0201-2745

TMS: https://orcid.org/0000-0002-0881-5808

GJS: https://orcid.org/0000-0002-8654-9867

AFT: https://orcid.org/0000-0002-9883-6134

FE: https://orcid.org/0000-0001-8528-0301

Keywords: silver; nanoparticles; polyol; automated synthesis; SAXS; DLS

**Abstract**

Metal nanoparticles have a substantial impact across different fields of science, such as photochemistry, energy conversion, and medicine. Among the commonly used nanoparticles, silver nanoparticles are of special interest due to their antibacterial properties and applications in sensing and catalysis. However, many of the methods used to synthesize silver nanoparticles often do not result in well-defined products, the main obstacles being high polydispersity or a lack of particle size tunability. We describe an automated approach to on-demand synthesis of adjustable particles with mean radii of 3 and 5 nm using the polyol route. The polyol process is a promising route for silver nanoparticles *e.g.*, to be used as reference materials. We characterised the as-synthesized nanoparticles using small-angle X-ray scattering, dynamic light scattering and further methods, showing that automated synthesis can yield colloids with reproducible and tuneable properties.

**Introduction**

Metal nanoparticles (in the following abbreviated to NPs) find many applications in medicine and technology[1,2]. Generally, synthesis routes of NPs strive to achieve a high level of control over size, shape, polydispersity and at the same time try to limit the extent of aggregation or agglomeration. In addition, the long-term stability and the rheological properties of such dispersions are of importance[2–5]. Problems with reproducible particle synthesis and colloidal stability have limited the availability of nanoparticle-based reference materials to comparatively few, despite of repeatedly demanded provision of nanoscale reference materials for environmental, health and safety measurements for many years[6]. The physicochemical aspects of NPs constitute complex requirements, determining the usability for a particular application. The various NP systems, often conceptually related to each other, exhibit different and unique challenges, when it comes to their synthesis, and hence the expected results. In this regard, silver NPs (Ag NPs) are difficult to synthesize, especially if tunability, long-term stability, and reproducibility are required[4,7,8]. This contrasts with Au NPs, for instance, which have recently become a "poster child" for a controllable and straightforward synthesis[9], a distinct size regulation and the realization of predicted nanoparticle sizes is state of the art[10]. Nevertheless, Ag NPs are in high demand due to their well-known anti-bacterial properties and their use in catalysis, photochemistry, sensing, and optoelectronics[3,4]. Therefore, synthesis routes for Ag NPs that deliver spherical particles of less than 20 nm in diameter and low polydispersity are of importance.

One such promising route, described by Kästner and Thünemann[11], is based on a polyol process published by Hu *et al.*[8]. In this synthesis, Ag NPs are formed from the reduction of Ag$^+$ ions in the presence of polyacrylic acid (PAA) in hot ethylene glycol (EG), where EG provides both the solvent and the reducing agent. This synthesis route delivers spherical NPs with a mean radius of 3.0 nm and a radii distribution width of 0.6 nm[12]. The nanoparticles are stabilized in an aqueous solution by a charged PAA shell and remain unchanged in a suspension for over six months. These properties suggest that such Ag NPs could constitute potential reference materials (RMs) for the quantification of the size distribution and concentration of nanoparticles[12]. RMs need to fulfil the criterion of long-term stability[3,4], normally verified by time-consuming experiments using procedures standardized internationally by ISO GUIDE 35, for example[13]. These entail storing samples for a certain amount of time (at least six month) and checking for deterioration over that period. For nanoparticle samples dispersed in a liquid phase, this criterion is especially hard to achieve due to sedimentation, potential biological activity, and aggregation. In the case of RMs, long term stability could constitute a severe issue if adjustments to the synthesis are required. For example, modification of a particle size by a slight change in the procedure, could have adverse and unpredictable effects on the stability. Moreover, in practice RMs are often prepared in bulk quantities with the intention of long-term storage exceeding the period of six months, either by the supplier or customer. This creates a risk of an unintentional expiration of the material under variable and ill-defined storage conditions.

Faster development of reference materials and circumventing these problems could be enabled by rapid on demand synthesis by an automated platform. Such an approach should reduce the need for a long shelf-life. In addition, fast, automated, and controlled synthesis in small batches would allow for more targeted testing of physicochemical properties, and thereby faster convergence on desired properties. In a classical laboratory approach, it is far less time-consuming to create one large batch and perform a full characterization on this system, compared to preparing several small batches distributed over time. This is mainly because of the potential effort required to run multiple syntheses with the required precision and level of reproducibility *(i.e.* the human factor). Single, large batch synthesis is likely to be problematic if the sample stability is unknown, which could potentially lead to substantial amounts of expensive materials to be wasted if an out-of-specs product is produced. Also, smaller batches benefit from better mixing and less temperature gradients in the reaction solution potentially improving uniformity of acquired materials.

Here, we conceptualize and implement an automated synthesis of Ag NPs using the polyol route to produce colloidally stable silver. In pursuing the automated synthesis of nanoparticles, the capabilities of the "Chemputer" are deployed, for the first time, into the field of inorganic chemistry. The Chemputer is a modular, automated platform developed by the Cronin group for execution of multi-step, solution based organic synthesis, including purifications[14–17]. Liquids can be transferred across a backbone, constructed from HPLC selection valves and syringe pumps. The Chemputer operates in a batch mode, common laboratory devices, such as heaters and glassware like round bottom flasks, are connected to the backbone, forming reaction modules. Solutions can be manipulated in these modules, and as all operations are controlled through a software script, reproducibility among individual syntheses is high. Likewise, any adjustments of the synthesis conditions, if required, are straightforward to implement and are documented in the reaction log file and a code versioning system.

**Materials and Methods**

*Synthesis - general considerations*

The synthesis of the silver nanoparticles was performed using a Chemputer platform[14–17]. The specific hardware and software considerations are described in the next section.

All chemicals were used as received: $AgNO_3$ (PanReac, for analysis), ethylene glycol (PanReac, puriss), polyacrylic acid (Sigma-Aldrich, molar mas $M_w$ = 1800 g/mol, ca. 25 monomer units per polymer chain), NaOH (Sigma-Aldrich), deionised water was taken from in-house Milli-Q system. The synthesis was adapted from ref.[11]. Prior to synthesis, all glassware and PTFE-coated stirring bars were thoroughly cleaned with aqueous nitric acid (30% wt.) and rinsed with deionised water (DI, MilliQ, 18 MΩ). Unless stated otherwise, all steps were conducted in a fully automated manner using the Chemputer (Fig. 1). Before the start of the synthesis, the elements of the Chemputer's backbone (syringes, valves, and tubing) were cleaned with 5 mL of nitric acid (30% wt.) by pumping it through the transfer paths and the reactor (round bottom flask), followed by 6 transfers of water (5 mL for each transfer) to remove any remaining acid. We also note here that nitric acid should be disposed of into a separate waste container to avoid contact with organic solvents. The transfer paths were subsequently dried by pumping volumes of 5 mL of acetone three times, followed by pumping 5 mL of dry air five times to remove acetone.

In our experiments, we had target radii of ~3 and 5 nm for which we used silver nitrate precursor concentrations of 33.4 mg/mL and 136.6 mg/mL, respectively. The specific reaction conditions are summarised in Table 1. A solution of polyacrylic acid in ethylene glycol (12.5 mL) was transferred within ~120 s to a 3-necked flask equipped with a PTFE stirring bar. Under stirring (400 rpm), the solution was heated to 210°C. When the temperature was reached, a solution of $AgNO_3$ in ethylene glycol (2.5 mL, for concentration see Table 1) was, due to its viscosity, slowly aspirated, transferred and then added rapidly (3 )s to the heated solution together with air (7 mL) to ensure quantitative addition (accounting for the tubings' dead volume). After 5 minutes, the colour of the solution had turned amber, changing to a deep brown over the following minutes. 15 minutes after addition, the heating was stopped, and the solution was left to cool to room temperature (25 °C), as was measured by a Pt100 thermocouple. After ambient temperature was reached (~ 1 hour), water (34 mL) was added, and the stirring stopped. The nanoparticle synthesis on the Chemputer took 4 days per batch. It should be noted that the synthesis takes 5 h, while the rest of time was spent on decanting in 24-hour-intervals. To not occupy the platform with decanting, the contents of the flask were transferred to narrow beakers (50 mL, 4 cm diameter) and left to sediment. After 24 hours, the supernatant was manually removed, the beaker refilled with 34 mL of DI water, and the sediment was redispersed. This routine was repeated two times, and after decanting for the third time we refilled the beaker with only little water (6.4 mL). Finally, an aqueous solution of NaOH (0.4 mL, 1% wt.) was added. After mixing, the opaque, brown suspension cleared to a dark, greenish-black solution.

The as-obtained nanoparticles were further characterized by transmission electron microscopy, X-ray and light scattering methods, as described below.

*Table 1. Synthesis conditions and values of the radii for the Ag nanoparticles.*

| Experiment code | $c_{AgNO_3}$ [mg/mL] | $V_{Ag, theoretical}$ [µL] | $c_{PAA}$ [mg/mL] | Target radius [nm] |
|---|---|---|---|---|
| NP3_I    |       |       |      |   |
| NP3_II   | 33.4  | 5.06  |      | 3 |
| NP3_III  |       |       |      |   |
| NP3_IV   |       |       | 53.8 |   |
| NP5_I    |       |       |      |   |
| NP5_II   | 133.6 | 20.21 |      | 5 |
| NP5_III  |       |       |      |   |

*Synthesis – software and hardware implementations*

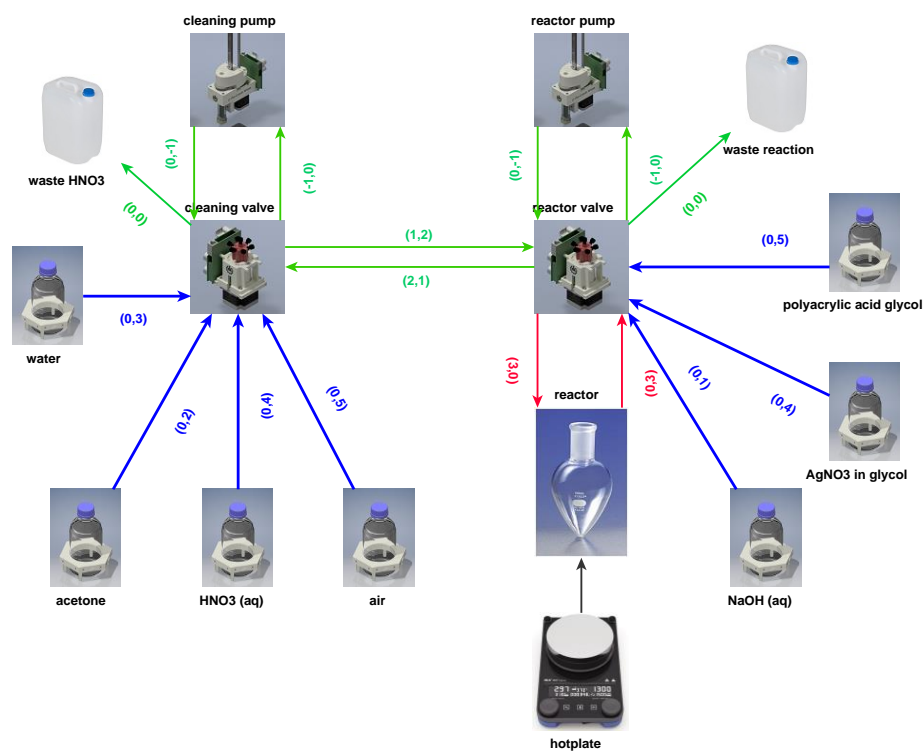

*Fig. 1. Graph scheme of the Chemputer platform employed for nanoparticle synthesis rendered from a GraphML[18] file (SI: reaction_graph_AgNP.graphml). Icons/pictures represent graph nodes and contain metadata for the computer control and available volumes. All these metadata are required by the API (the "Chempiler"[19]). Arrows represent connections; the black arrow between the hotplate and the reactor a temperature control connection, all other arrows represent tubing with metadata such as internal volume and the connection information in parenthesis. Directions of arrows encode possible liquid movement directions. Green arrows represent tubing in the Chemputer backbone, the blue connections tubing to storage flasks and the red ones tubing to reaction vessels. Note that arrow colours are only included for clarity and are not relevant to the automation platform and code, also note that two arrows pointing in opposite directions encodes one piece of tubing with no restrictions on flow direction. For more in-depth discussion of the graph and its software implementation into the Chempiler, refer to[16,19].*

The nanoparticle synthesis diagram required by the Chemputer is presented in Fig. 1. The diagram is directly rendered from a GraphML[18] file (XML-based, see SI: file included). The file represents all the reactant solutions, glassware, devices etc. and their topology/connections indicated by arrows. Each object depicted in Fig. 1 contains metadata, which are read by the Python API (the "Chempiler")[16] to run the Chemputer. The synthesis conditions are defined in the previous section and Table 1, based on which we present their implementation in the code (steps I-VI). Below, we describe the consecutive functional steps, while the actual Python code snippets are shown in the SI.

Step I) Before the synthesis, tubing connecting reservoirs and pumps must be filled with reagents, to guarantee reproducible volume transfers. Potential glycol and acetone residues are then washed away with water thoroughly, followed by washing the backbone tubing with nitric acid (30%). Nitric acid residues are washed away with water, water is washed away with acetone and acetone is dried of with air transfers (SI: Snippet 1).

Step II) The as-prepared platform can be used for synthesis. First transferring polyacrylic acid in glycol to the reactor, followed by transferring air to empty the tubing path. The backbone is cleaned again as described before, heating and stirring of the reactor is started, and once the temperature is reached, the solution of silver nitrate in glycol is added (SI: Snippet 2).

Step III) To ensure quick and complete addition in one portion, the reactor syringe is charged with the solution and additional 7 mL of air as headspace. The syringe then is discharged into the reactor at maximum speed (3 s for the 2.5 mL of solution) (SI: Snippet 3).

Step IV) After a waiting time of 15 minutes, the heating is stopped, and the reactor left for cooling to room temperature (SI: Snippet 4).

Step V) Upon reaching room temperature, a specific volume of water is added, and the same volume removed after 24 h, decanting facilitated by adjusting the level of the tubing end(SI: Snippet 5).

Step VI) The resulting suspension is diluted with water and the pH manually adjusted by adding NaOH (SI: Snippet 6).

*Characterisation*

Small- and wide-angle X-ray scattering (SAXS/WAXS) measurements were conducted using the MOUSE (Methodology Optimization for Ultrafine Structure Exploration) instrument[20]. X-rays were generated from a microfocus X-ray tube, followed by multilayer optics to parallelize and monochromatize the X-ray beams to a wavelength of Cu K$\alpha$ ($\lambda$ = 0.154 nm). Scattered radiation was detected on an in vacuum Eiger 1M detector (Dectris, Switzerland), which was placed at multiple distances between 137 and 2507 mm from the sample. Samples and backgrounds were measured in flow-through capillaries. The resulting data were processed and scaled to absolute intensity using the DAWN software package in a standardized complete 2D correction pipeline with uncertainty propagation[21–23]. The data was fitted and analysed using the program McSAS, a Monte Carlo method for fitting SAXS data[24,25] (Table 2, and SI). We used the assumption that the Ag particles were spherical in shape, and the scattering length densities (SLDs) used for the fits were $SLD_{water}$ = 9.4691·10$^{-6}$ Å$^{-2}$, $SLD_{Ag}$ = 7.7854·10$^{-5}$ Å$^{-2}$. As the intensity is expressed in absolute units, the resulting size distributions are also absolute in terms of calculated volume fractions.

Dynamic Light scattering (DLS) was measured on a Malvern Instruments Zetasizer. Sample solution (10 µL) was diluted with an aqueous solution of NaOH (pH 10) to 1 mL. Measurement was performed at 25 °C in ZEN0040 disposable cuvettes (Malvern Instruments) after an equilibration time of 120 s under a scattering angle of 173° (backscatter condition), automatic measurement duration, 5 measurements and 1 s inter measurement delay time. The data were collected and automatically processed using Zetasizer Software version 8.

For TEM measurements, 1 mL of sample, directly after synthesis and before pH adjustment was sedimented with a small benchtop centrifuge (a = 100 G), 0.9 mL of the supernatant was removed, 0.9 mL of water was added, and the sediment was redispersed. This was repeated for 7 times to remove ethylene glycol and other solutes. The sample was diluted with 4 mL of Milli-Q water, 10 µL added onto the copper grid (3 mm holey carbon-coated Cu-grid (Lacey Carbon, 400 mesh)) and after drying, cleaned in a plasma cleaner for 15 s. Bright field electron microscopy (TEM) images were obtained on a Talos F200S Microscope (Thermo Fisher Scientific) operating at 200 kV equipped with a Ceta 16M camera.

Powder X-ray diffraction (XRD) patterns from dried NP3_I and NP5_I samples were collected on a D8 Bruker Diffractometer equipped with a LYNXEYE XE-T detector operating in a Bragg-Brentano geometry (reflection mode). Small volumes (~0.5 mL) of the colloidal suspensions of Ag NPs were deposited onto low-background Si holders so that the liquid was covering an entire surface of the 1" wafer. The sample holders were put into an oven at 35°C and left to dry for 2 hours. Diffraction was measured with Cu Kα radiation (1.5406 Å, 40 kV and 40 mA) from 5-100° using a step size 0.010° (2Θ) and a scanning time of 384 s per step. The diffraction profiles were deconvoluted by fitting pseudo-Voigt functions from which the peak width/broadening values (full width at half-maximum, FWHM) were obtained. The instrumental angle-dependent peak broadening was determined by measuring a corundum standard. The crystallite sizes were calculated from the fit peak FWHM values corrected for the instrumental broadening using the Scherrer formula.

*Table 2. Summary of the particle characteristics derived from SAXS and DLS. The mean radius, R; and its radii distribution width, σ; volume fraction, ϕ; concentration, c, were all derived from SAXS. Uncertainty of ϕ values is about 10%. The hydrodynamic radii, $R_h$; and the $\sigma_h$ radii distribution width were obtained from DLS (volume-weighted size distribution).*

| Experiment code | R (nm) | σ (nm) | ϕ | c (mg/mL) | $R_h$ (nm) | $\sigma_h$ (nm) | Yield |
|---|---|---|---|---|---|---|---|
| NP3_I   | 3.34 ± 0.01 | 0.81 ± 0.20 | $2.0 \cdot 10^{-4}$ | 2.09 ± 0.054 | 7.59 ± 0.83  | 2.33 ± 0.73 | 0.3153 |
| NP3_II  | 3.37 ± 0.01 | 0.83 ± 0.21 | $2.2 \cdot 10^{-4}$ | 2.30 ± 0.060 | 8.80 ± 0.92  | 2.92 ± 0.83 | 0.3463 |
| NP3_III | 3.36 ± 0.01 | 0.85 ± 0.25 | $2.2 \cdot 10^{-4}$ | 2.35 ± 0.061 | 7.82 ± 0.64  | 2.55 ± 0.49 | 0.3547 |
| NP3_IV  | 3.52 ± 0.01 | 0.99 ± 0.25 | $2.1 \cdot 10^{-4}$ | 2.18 ± 0.057 | 6.91 ± 0.83  | 1.89 ± 0.52 | 0.3286 |
| NP5_I   | 5.26 ± 0.01 | 1.33 ± 0.17 | $1.2 \cdot 10^{-3}$ | 12.50 ± 0.325 | 11.39 ± 0.40 | 3.99 ± 0.38 | 0.4709 |
| NP5_II  | 5.35 ± 0.01 | 1.65 ± 0.13 | $1.1 \cdot 10^{-3}$ | 10.99 ± 0.29 | 12.75 ± 0.69 | 5.44 ± 0.49 | 0.4147 |
| NP5_III | 5.33 ± 0.01 | 2.13 ± 0.35 | $1.4 \cdot 10^{-3}$ | 14.27 ± 0.37 | 15.79 ± 1.31 | 9.05 ± 1.25 | 0.5381 |

**Results and Discussion**

Our first automated experiment aimed at producing NPs of ~3 nm in radius, by following the synthetic route and conditions reported earlier[11]. We evaluated the overall reproducibility by running the synthesis in quadruplicate (see Tables 1 and 2: NP3: I - IV) and performing the characterisation of the NPs with scattering methods. Fig. 2 shows the SAXS curves from four runs and the particle size distributions derived from the Monte Carlo fits. The size-distributions were obtained under the assumption that particles were represented by simple spherical form factors, which we based on TEM images (see SI: Fig. S1). In this regard TEM imaging does not provide sufficiently good statistics to extract the size distributions of Ag NPs, but is necessary to confirm their shape, which is an important assumption for SAXS data fitting. The scattering patterns from all four runs exhibit a very small spread within the experimental uncertainties (Figs. 2A-D). This clearly indicates that the statistically significant differences among the samples are minor (see the overlapping curves from NP3-series in Fig. 2H), which confirms a high reproducibility of the process. Overall, based on the shape of the profiles we can conclude that the particles were relatively monodisperse, and unaggregated. Alternatively, we also considered the core-shell spheres as a form factor (Ag and PAA, respectively), but due to the $\Delta SLD^2$(Ag, PAA) >> 1000 the core-shell form factor was not measurable in SAXS. The as-derived volume-weighted size distributions all have a mean value of ~3.3 nm and a standard deviation of the distribution of ~0.8 nm, where the total volume fraction of particles is ~0.2% (Table 2; complete reports generated by the McSAS are included in the SI: Supporting Files). The distributions are overlaid in SI: Fig. S2 for a better comparison.

In the second series of experiments (Tables 1 & 2: NP5: I-III) our goal was to modify the synthesis so that ~5 nm NPs would be obtained. We chose the value of 5 nm to check how accurately we could control the reaction. Under the assumption that all the other reaction conditions were kept constant, the ~(5/3) increase in radius implied the cubic increase in the required concentration of silver (Table 1: ~4.2 times). Please note that these were approximate conditions since we had not known the size distributions of our NP3 series *a priori*. Under these assumptions, we performed three syntheses, and the resulting scattering curves are shown in Figs. 2 E-G. As in the case of NP3, the resulting NPs were unaggregated, but exhibited relatively higher polydispersity (SI: Fig. S2). On average the particle mean radius was ~5.3 nm, but a standard deviation of the distribution ranged from ~1.3 - 2.1 nm, depending on a run (Table 2 and Fig. 2: NP5). It is important to note, that both the NP3 (Fig. 2A-D) and NP5 series SAXS curves (Fig. 2E-G) exhibit upturns in the profiles for $q < 0.04$ nm$^{-1}$, where the data points have relatively small uncertainties for the NP5 series. These upturns might therefore superficially point to the presence of very minor populations of larger species or aggregation. However, if we compare the measured intensities against the low-q instrumental background level of our SAXS instrument (see Fig. 4 in[20]), it turns out that the upturn is expected, due to the instrumental background at these levels of intensity.

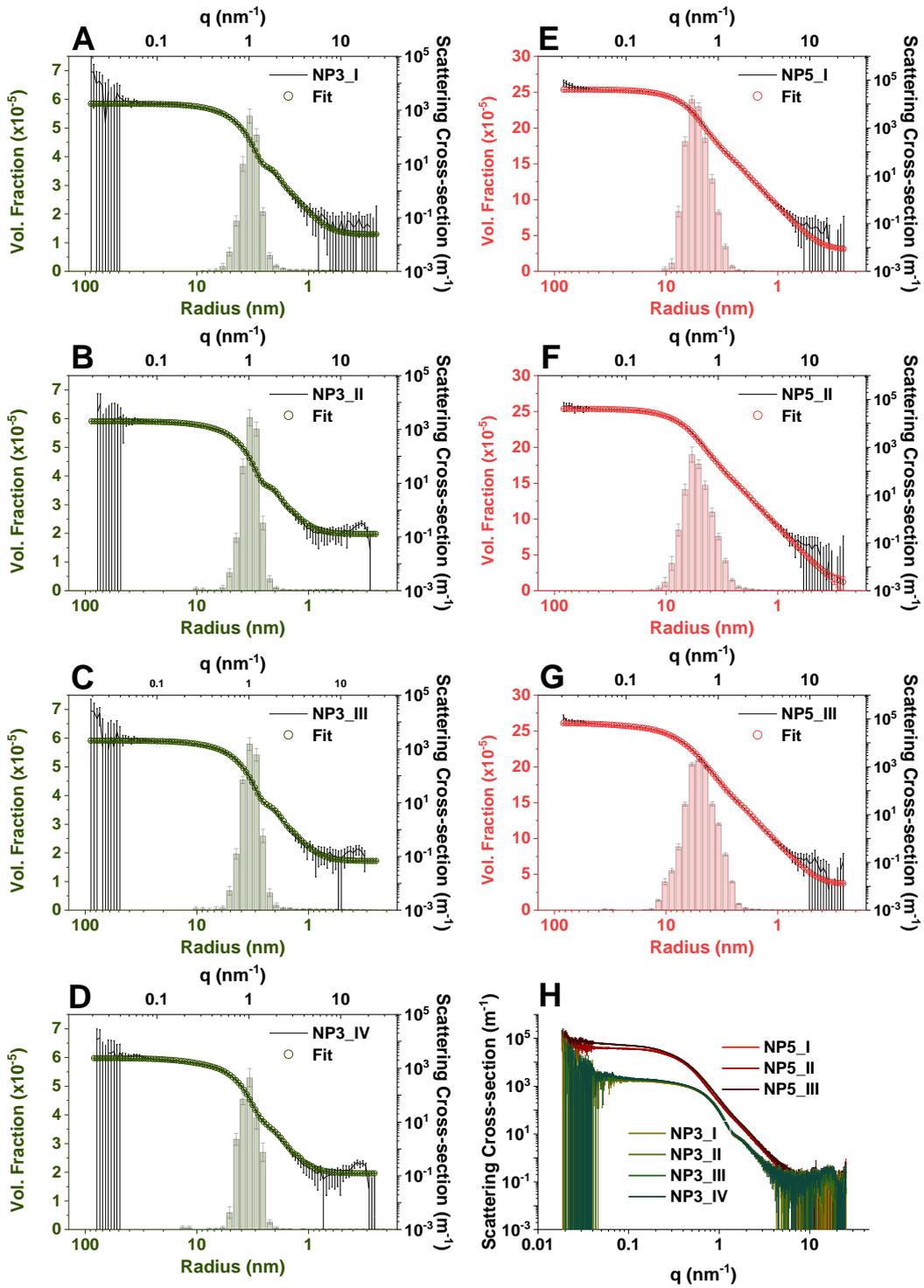

*Fig. 2. SAXS scattering profiles from NP3 and NP5 (see also Tables 1 and 2). Scattering profiles and selected fits using a polydisperse sphere model, where the corresponding size distributions were derived using a Monte Carlo method implemented in McSAS; A-D: NP3_I-IV; E-G: NP5_I-III; H: comparison of all the profiles from NP3 and NP5 series on an absolute intensity scale.*

All the particles from NP3 and NP5 series were also characterised using DLS (Table 2, Fig. 3; SI: Fig. S3). The as-derived volume-weighted radii are an average from, typically, >10 measurements collected over several days (up to 6 days) on samples extracted from each batch. The DLS data, in general, correlate with those from SAXS, but the measured radii from light scattering are systematically larger than those measured by the X-rays, by a factor of >2. DLS measures hydrodynamic size of particles, which in our case includes the size of Ag NPs as well as the PAA shell which provides colloidal stability. Hence, the DLS measurements indicate that we deal with core-shell NPs of sorts. The absence of this shell in SAXS and TEM (SI: Fig. S1) indicates that it is possibly highly hydrated and hence of very low electron density and not persistent upon drying and in case of TEM, plasma cleaning (see Methods). Arguably, depending on a point of view, such a PAA shell may not be considered a shell (in the sense of core-shell NPs), as it has a possible liquid-like dynamic character and is not an integral part of the NPs. This PAA coating, however, is crucial for providing long-term colloidal stability of the NPs[8,11]. Regardless of the physicochemical classification of the "shell" we can still evaluate the thickness of the PAA layer at the surface of the NPs, from the juxtaposition of the mean values of the hydrodynamic radii from DLS and NPs' radii from SAXS (Table 2). In the case of the NP3 series the PAA layer was 4.4 ± 0.8 nm thick, while for the NP5 series it was 8.0 ± 2.2 nm. The measured values of the shell thickness for the NP3 series are lower than those reported for the analogous series in ref.[11], where the layer was found to be ~7 nm. However, considering the widths of the size distributions (*i.e.* standard deviations in Table 2) the difference is quite insignificant.

Two selected samples from each series, NP3_I and NP5_I were also characterised with XRD. In this case the samples were deposited and dried on silicon wafer holders and formed thick films. The diffraction patterns are shown in the SI (Fig. S4). The XRD confirmed that the NPs were crystalline, and that elemental Ag was the only crystalline phase present. The size of the NPs is clearly manifested by peak broadening (FWHM), where 5.3 nm NPs yield narrower reflections in comparison with the 3.3 nm counterparts (see for instance 111 peak in SI: Fig. S4). From the fitted peak profiles we evaluated crystallite size, which were 2.2 ± 1.0 nm for NP3_I, and 5.6 ± 3.8 nm for NP5_I. These values correctly represent the expected trends and agree within the uncertainties with the SAXS data. This indicates that the NPs are single-crystalline in nature.

In overall, from the presented measurements it is evident that the automated synthesis with the Chemputer delivers NPs of expected specifications in terms of size distributions. Since all the SAXS measurements are scaled to absolute units, the resulting volume-weighted size distributions are also absolute. Hence, the total volume fractions of Ag NPs can be easily expressed as a concentration of silver in the analysed samples (Table 2). By finding the ratio of these measured concentration values and the initial reactant concentrations (Table 1), we calculated the overall yields of each run (Table 2). Based on these data, one can see that for NP3 series the yields were from 30% to 35%, while for the NP5 series the values were higher ranging from 41% to 53%. The spreads of yields within and the different ranges of yields between the series, resulted from the decanting procedure (see Methods: Synthesis – software and hardware implementations, Step 5; SI: Snippet 5). This step involved removal of the supernatant from above of the sedimented suspension of the NPs. Although it was performed with the Chemputer, it involved estimation of the position of the interface between the two phases. The simplest approach to automated decanting was to adjust the PTFE tubing to be suspended above the level of sedimented particles (~5 mm). For determining the tubing level, at least one synthesis should be performed and the tubing visually adjusted. The lack of feedback on potential particle removal was, however, problematic as is evident from the values of yield. Approaches more amenable to automation like ultrafiltration could be included, potentially improving this purification step. Moreover, the automated titration with NaOH could be possible, however the Chemputer is tailored towards larger volumes >> 1 mL and working with smaller volumes would require further modification of the setup. We must point out, however, that the described above limitation is not so much relevant for the automated synthesis of reference materials in which the size distribution and long-term stability are the major quality criteria.

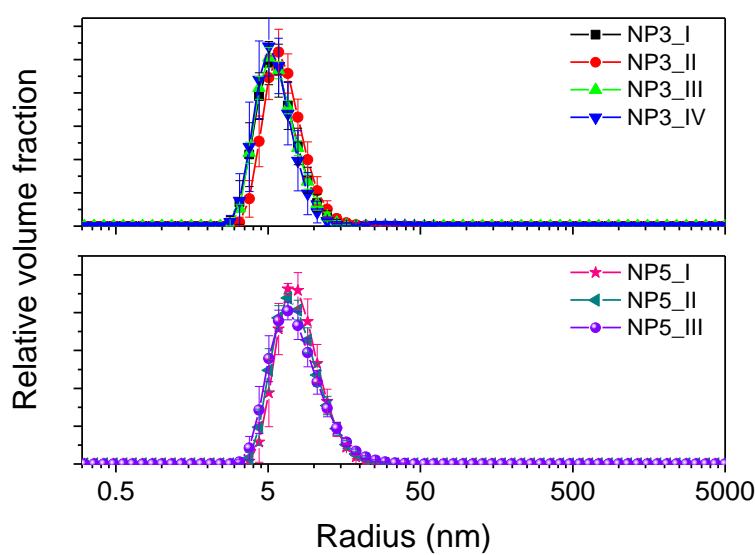

*Fig. 3. Volume-weighted size distributions for the NP3 (upper) and NP5 (lower) series derived from the DLS measurements. Each presented data set is an average of typically 10 individual measurements recorded over a period of 6 days. The mean and standard deviation values are reported in Table 2. The corresponding intensity-weighted distributions can be found in the SI (Fig. S3).*

**Outlook**

The used process employs a polyol, such as ethylene glycol, as solvent and reductant at the same time. By heating the polyol and injecting silver ions dissolved in ethylene glycol, polyol is oxidized to aldehydes and carboxylic acids, reducing the silver ions to $Ag^0$ in turn[2,4,11]. Parameters such as the reaction time, polyol viscosity or temperature are variables that control synthesis and the final properties of Ag NP, most notably, shape, size, and their distributions[2,5]. For instance, an increase in reaction time also broadens the size distribution of the particles, which essentially is not desirable. Similarly, another possibility is to tune the viscosity, by either employing a more viscous polyol or by polymeric additives[3]. The herein employed synthesis used polyacrylic acid of defined molecular weight as a polymeric additive, which also stabilizes the as-obtained colloid by charged surface capping the particles. Viscosity tuning, however, requires changing reactants, which may affect the mechanisms of NP formation, and in the case of automated synthesis requires different stock solutions and adjustment of synthetic procedures for the custom target.

In our approach we followed the hypothesis that size is controlled by availability of silver ions. Under this assumption, particle size follows a cubic relationship to the amount of reduced silver. This is approximated, because of the possible reverse oxidation of silver from $Ag^0$. Therefore, under this assumption, *e.g.* to double the radius of the NPs, the dissolved silver concentration would need to be increased eight-fold. In experimental terms, we injected 2.5 mL of Ag-bearing EG solution into 12.5 mL of hot EG with PAA (see Methods), hence the increase in silver concentration could be effectively achieved either by increasing the concentration of the silver solution, or, as a compromise, by increasing the volume of this solution. Both approaches have their advantages and disadvantages. On the one hand an increase in the concentration may be limited by the solubility of silver nitrate in EG, but at the same time it reduces the need for the direct addition of large volumes of silver-bearing solutions, which are added to the large volume of hot EG. Larger addition volumes could negatively impact the stability of temperature (210°C) of the reaction solution, broadening the time range of seed creation, effecting the resulting NP size distribution. On the other hand, the increase in silver concentration (*i.e.,* a fixed volume) also increases the kinetic rates of the reaction and consequently increases the nucleation and growth rates. As a result, the NPs are larger, but they also exhibit broader size distributions, in comparison with their smaller counterparts. This is indeed what we observed for the NP3 and NP5 series (Tables 1 and 2, Figs. 2 and 3) in which the concentration of silver was a variable. Hence, for either of the strategies, we would expect a broader size distribution for larger NPs, but the extent of this effect may differ.

**Conclusions**

In conclusion, we described an automated synthesis of silver nanoparticles to produce adjustable monodisperse particles. The nanoparticles where thoroughly characterized by SAXS, DLS, TEM and XRD. The approach is an important first step towards the automation of nanoparticle syntheses in a modular, multipurpose automation platform. The modularity of the Chemputer opens many possibilities for the synthesis of a variety of different nanoparticle morphologies and sizes and potentially more complex structures.

**Acknowledgements**

We thank Paul Andrle for help with the synthesis and preparation of stock solutions, Oskar Haase and Ralf Bienert for initial characterisations of the particles with SAXS, Carsten Prinz for TEM measurements and Christoph Naese for quick and carful construction of numerous parts involved in the project.

**Supplementary Information:** Supplementary files: (1) GraphML for the Chemputer, (2) 7 SAXS fitting reports from MCSAS; Supplementary figures: Figs. S1-S4; Supplementary computer code: Snippets 1-6

# Supplementary Information

## for

***Towards automation of the polyol process for the synthesis of silver nanoparticles***

Jakob Wolf[1,2], Tomasz M. Stawski[1]\*, Glen J. Smales[1], Andreas F. Thünemann[1],

Franziska Emmerling[1,3]\*\*

\* tomasz.stawski@bam.de; \*\* franziska.emmerling@bam.de

[1]Federal Institute for Materials Research and Testing (BAM), 12489, Berlin, Germany;

[2]Max Planck Institute of Colloids and Interfaces, 14476 Potsdam, Germany;

[3]Humboldt-Universität zu Berlin, Department of Chemistry, 12489, Berlin, Germany

## *Supplementary files*

*All the supporting files not, which are not included in this pre-print can be downloaded from here: https://doi.org/10.5281/zenodo.5910614*

1. The graphml file: **reaction_graph_AgNP.graphml** is included. It contains topological information (Fig. 1 in the main text) about the reaction setup and metadata with reaction condtions. It used by the Python API used to control the Chemputer.

2. SAXS reports. The complete report sheets generated by McSAS are included. They contain extended information characterising the size distributions and the fitting parameters.

   NP3_I: **saxs_report_NP3_I.pdf**

   NP3_II: **saxs_report_NP3_II.pdf**

   NP3_III: **saxs_report_NP3_III.pdf**

   NP3_IV: **saxs_report_NP3_IV.pdf**

   NP5_I: **saxs_report_NP5_I.pdf**

   NP5_II: **saxs_report_NP5_II.pdf**

   NP5_III: **saxs_report_NP5_III.pdf**

***Supplementary figures:***

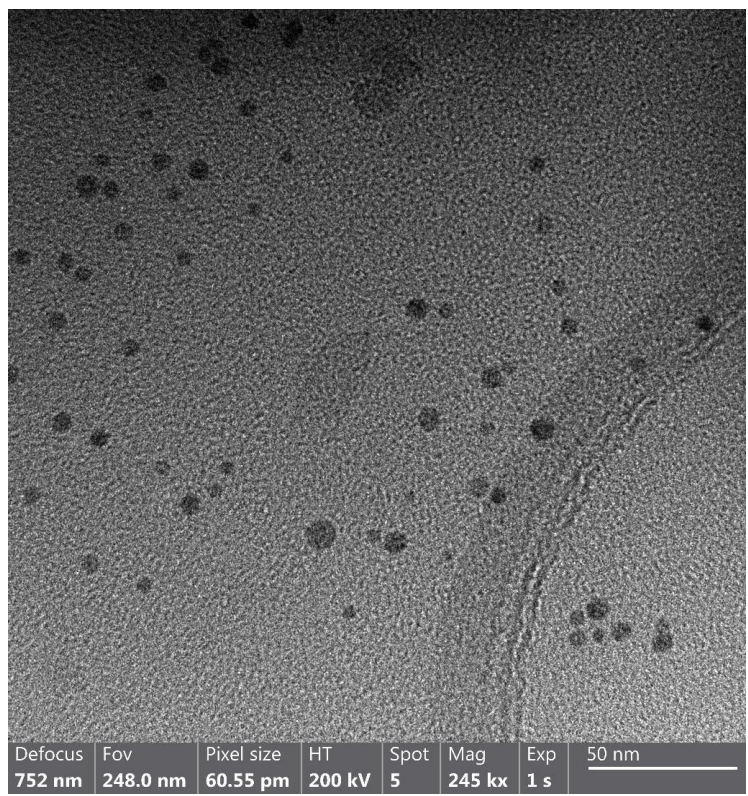

Fig. S1. Bright-field TEM image of the NP3 series nanoparticles. It shows that the NPs were indeed spherical and relatively monodisperse.

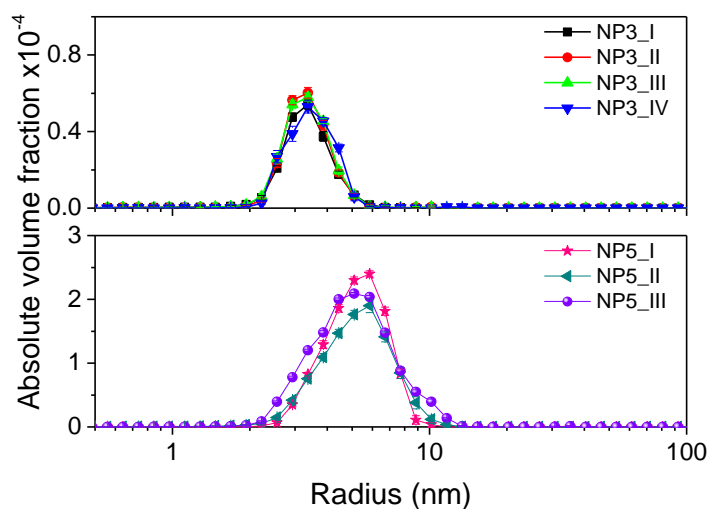

Fig. S2. Volume-weighted size distributions for the NP3 (upper) and NP5 (lower) series derived from the SAXS measurements. The mean and standard deviation values are reported in Table 2 in the main text.

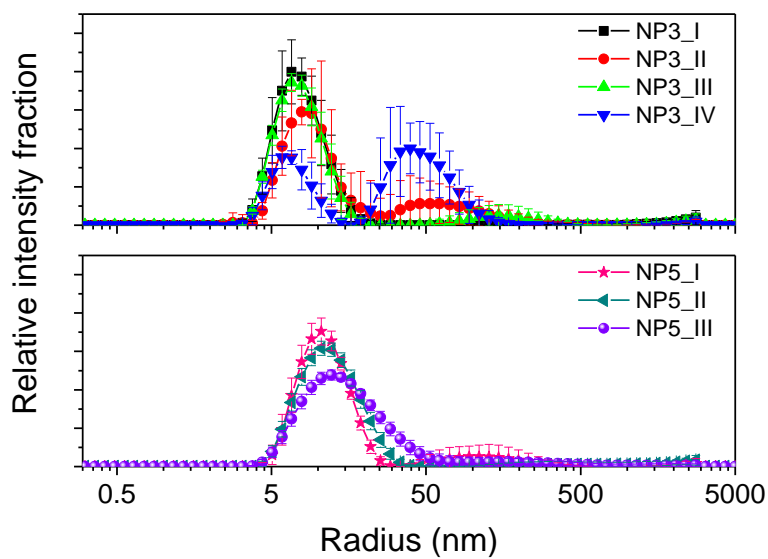

Fig. S3. Intensity-weighted size distributions for the NP3 (upper) and NP5 (lower) series derived from the DLS measurements.

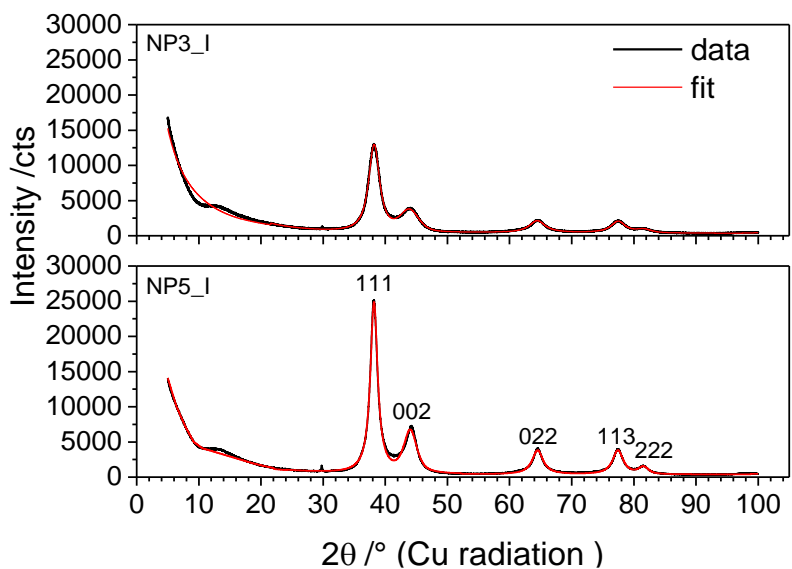

Fig. S4. XRD from dried NP3_I (upper) and NP5_I (lower) samples. The black curves represent measured data and the red ones fitted profiles. Both diffraction patterns unequivocally correspond to single-phase Ag (cubic, F m -3 m). Miller indices of the diffraction peaks are provided.

## Supplementary computer code

**Snippet 1.** Cleaning and priming of the Chemputer.

```
chempiler.move('AgNO3 in glycol', 'waste reaction', priming_volume,
initial_pump_speed=default_speed_viscous)
# priming_volume = 1.5 mL, default_speed_viscous = 5 mL/min
chempiler.move('polyacrylic acid in glycol', 'waste reaction', priming_volume,
initial_pump_speed=default_speed_viscous)

# make sure glycol is washed away for chemical compatibility
for i in range(3):
    chempiler.move('water', 'waste reaction', default_cleaning_volume)
# default_cleaning_volume = 5 mL

for i in range(3):
    chempiler.move('HNO3 (aq)', 'cleaning pump', default_cleaning_volume)
    chempiler.move('cleaning pump', 'waste HNO3', default_cleaning_volume)

for i in range(3):
    chempiler.move('water', 'cleaning pump', default_cleaning_volume)
    chempiler.move('cleaning pump', 'waste HNO3', default_cleaning_volume)

chempiler.move('water', 'waste reaction', default_cleaning_volume, repeats=3)
chempiler.move('acetone', 'waste reaction', default_cleaning_volume, repeats=3)
chempiler.move('air', 'waste reaction', default_cleaning_volume, repeats=3)
```

**Snippet 2.** Synthesis initialisation.

```
chempiler.move('polyacrylic acid in glycol', 'reactor', 12.5,
        initial_pump_speed=default_speed_viscous)

# ensure quantitative transfer
chempiler.move('air', 'reactor', 5)

# clean up
chempiler.move('water', 'waste reaction', 5, repeats=3)
chempiler.move('acetone', 'waste reaction', 5, repeats=3)
chempiler.move('air', 'waste reaction', 5, repeats=3)

# heat reactor
```

```python
chempiler.stirrer.set_temp('hotplate', 210)
chempiler.stirrer.set_stir_rate('hotplate', default_stirring_rate)
# default stirring rate = 400 rpm
chempiler.stirrer.stir('hotplate')
chempiler.stirrer.heat('hotplate')

# make sure tubes are dry
chempiler.move('air', 'reactor', 10)
chempiler.move('air', 'waste reaction', 40)

# wait for reaching the reduction temperature
chempiler.stirrer.wait_for_temp('hotplate')
```

**Snippet 3.** Solution injection.

```python
chempiler.move('AgNO3 in glycol', 'reactor pump', 2.5,
initial_pump_speed=default_speed_viscous)
chempiler.move('air', 'reactor pump', 7)
chempiler.move('reactor pump', 'reactor', 9.5, end_pump_speed=50) #this has to
happen rapidly
```

**Snippet 4.** Stopping the reaction.

```python
# wait for reduction
chempiler.wait(reduction_time)

# wait for cooldown
chempiler.stirrer.set_temp('hotplate', room_temperature)
# room_temperature = 25 °C
chempiler.stirrer.wait_for_temp('hotplate')
```

**Snippet 5.** Decantation procedure.

```python
# automated decanting
for i in range(3):
    chempiler.stirrer.stir('hotplate')
    chempiler.move('water', 'reactor', 34)
    chempiler.stirrer.stop_stir('hotplate')
    chempiler.wait(24*one_hour)
    chempiler.move('reactor', 'waste reaction', 35)
```

**Snippet 6.** Final steps.

```
# automated dilution
chempiler.stirrer.stir('hotplate')
chempiler.move('water', 'reactor', 6.4)
```